\documentclass[pre,twocolumn,preprintnumbers,amsmath,amssymb]{revtex4}

\usepackage{graphicx}
\usepackage{dcolumn}
\usepackage{bm}
\usepackage{parskip}
\usepackage{color}
\usepackage{subfigure}
\usepackage{amssymb}
\usepackage{amsmath}
\usepackage{amssymb}
\usepackage{graphicx}

\def\be{\begin{eqnarray}}
\def\ee{\end{eqnarray}}
\def\be{\begin{equation}}
\def\ee{\end{equation}}
\begin{document}

\title{Multiparameter universality and  conformal field theory \\for anisotropic confined systems: test by Monte  Carlo simulations
}

\author{Volker Dohm, Stefan Wessel, Benedikt Kalthoff, and Walter Selke}

\affiliation{Institute for Theoretical Physics, RWTH Aachen University, 52056 Aachen, Germany}

\date {June 17, 2021}

\begin{abstract}
Analytic predictions have been derived recently by V. Dohm and S. Wessel,
Phys. Rev. Lett. {\bf 126}, 060601 (2021) from anisotropic $\varphi^4$ theory and conformal field theory for the amplitude ${\cal F}_c$ of the critical free energy of finite anisotropic systems in the two-dimensional Ising universality class. These predictions employ the hypothesis of multiparameter universality. We test these predictions by means of high-precision Monte Carlo (MC) simulations for ${\cal F}_c$ of the Ising model on a square lattice with isotropic ferromagnetic couplings between nearest neighbors and with an anisotropic coupling between next-nearest neighbors along one diagonal. We find remarkable agreement between the MC data and the analytical prediction. This agreement supports the validity of multiparameter universality and invalidates two-scale-factor universality as ${\cal F}_c$ is found to exhibit a nonuniversal dependence on the microscopic couplings of the scalar $\varphi^4$ model and the Ising model.
Our results are compared with the exact result for ${\cal F}_c$ in the three-dimensional $\varphi^4$ model with a planar anisotropy in the spherical limit. The critical Casimir amplitude is briefly discussed.
\end{abstract}
\maketitle
%
%
%
\renewcommand{\thesection}{\Roman{section}}
\renewcommand{\theequation}{1.\arabic{equation}}
\setcounter{equation}{0}
\section{ Introduction}
A widely-held belief in the traditional theory of critical phenomena has been the general validity of two-scale-factor universality for the singular behavior of confined thermodynamic systems with short-range interactions \cite{cardy1987,cardy1988,privman1990,priv,pri,brankov,pelissetto,henkel}. The most fundamental
quantity is the free-energy density $f$.
For $d$-dimensional systems with a characteristic length $L$
the asymptotic critical behavior of the singular part $f_s$ of $f$ (divided by $k_BT$) has been hypothesized to be described by the scaling form for $d<4$ \cite{pri}
\begin{eqnarray}
\label{freesing}
f_s (t, h, L) = L^{-d} \; F(C_1 t L^{1/\nu},
C_2 h L^{\beta \delta/\nu})
\end{eqnarray}
for large $L$, small $t=(T-T_c)/T_c$ and small ordering field $h$, with universal critical exponents $\nu, \beta, \delta$.
It was asserted that, for given geometry and boundary conditions,  the scaling function $ F(x, y)$ is universal, i.e., that the two metric factors $C_1$ and $C_2$ are the only nonuniversal parameters entering (\ref{freesing})\cite{cardy1987,cardy1988,privman1990,priv,pri,brankov,henkel,pelissetto}.
The universal structure of (\ref{freesing}) is referred to as two-scale-factor universality \cite{priv,pri}. It was believed that (\ref{freesing}) is valid for all systems in a universality class, i. e.,  for both isotropic and weakly anisotropic systems, since it was argued  that universality can be restored \cite{Indekeu}, reintroduced \cite{nightingale}, or repaired \cite{priv} in the latter systems by a suitable anisotropic scale transformation restoring asymptotic isotropy \cite{cardy1987,cardy1996,cardy1983}.
Correspondingly the same finite-size scaling function of the critical Casmir force has been predicted for anisotropic superconductors and isotropic superfluids  \cite{wil-1} and the absence of anisotropy effects  has not been questioned in a comment \cite{comment} on the prediction of \cite{wil-1} below $T_c$.
However,  it has been shown \cite{cd2004,dohm2006,dohm2008,kastening-dohm} that restoring isotropy does not eliminate nonuniversality since the transformed isotropic system still carries the nonuniversal anisotropy information of the original system both in its changed geometry and in the nonuniversal orientation of its transformed boundary conditions. In particular the prediction of \cite{wil-1} for anisotropic superconducting films has been refuted \cite{dohm2018}.

Recently a unified hypothesis of multiparameter universality \cite{dohm2018}, originally introduced for critical bulk amplitude relations \cite{dohm2008}, was formulated for bulk and confined anisotropic systems. Subsequently its exact validity was proven for the critical bulk order-parameter correlation function and for a critical bulk amplitude relation in two dimensions \cite{dohm2019}. For confined systems this hypothesis predicts that $f_s$ depends on $d(d+1)/2+1$ nonuniversal parameters. In particular, for a system of a cubic shape with the volume $L^d$ at $t=0, h=0$, it implies that the singular part of the total free energy at criticality
\begin{eqnarray}
\label{freeamplitude}
{\cal F}_c(q,\Omega)=L^d f_s (0,0,L)=F(0,0)
\end{eqnarray}
depends on $d(d+1)/2-1$ nonuniversal parameters, i.e. two nonuniversal anisotropy parameters in two dimensions.
More specifically, in the latter case these parameters are the angle $\Omega$ describing the orientation of the two principal axes and the ratio $q=\xi_{0\pm}^{(1)}/\xi_{0\pm}^{(2)}$ of the two principal correlation lengths
$\xi_\pm^{(\beta)}=\xi_{0\pm}^{(\beta)}|t|^{-\nu}, \beta=1,2$,
above $T_c$ $(+)$ and below $T_c$ $(-)$ with
$t=(T-T_c)/T_c$ where $\nu=1$ for Ising-like systems \cite{dohm2019}. If correct, this hypothesis has serious consequences for the predictability of the critical amplitude ${\cal F}_c(q,\Omega)$ and the ensuing critical Casimir amplitude. As pointed out earlier \cite{dohm2018,night1983},  contrary to the feature of two-scale-factor universality, the angle $\Omega$ generically depends in an unknown way on the anisotropic interactions, thus ${\cal F}_c(q,\Omega)$ remains to be an unpredictable amplitude for all anisotropic systems whose orientation of the principal axes is unknown. For the same reason the critical Casimir amplitude remains unpredictable for such anisotropic systems. This refutes the claim \cite{diehl2010} that the anisotropy encountered in weakly anisotropic systems is of a fairly harmless kind. It also refutes  common assertions (see, e.g., \cite{KrDi92a,toldin2013,hucht2011,vasilyev2009,diehl2014})  that the critical Casimir amplitude is independent of microscopic details and thus depends only on a few global and general properties,  such as the dimension $d$, the number of components of the order parameter, the shape of the confinement, and the type of the boundary conditions (BC).

\begin{figure}
\includegraphics[clip,width=6.0cm]{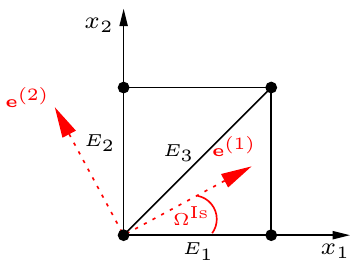}
 \vspace{0cm}
 \caption{(Color online)  Lattice points of the "triangular-lattice" Ising  model (\ref{IsingH}) on a square lattice with couplings $E_1,E_2,E_3$. Dotted arrows ${\bf e}^{(1)}$ and ${\bf e}^{(2)}$ denote the principal directions for  $E_1>E_2>0, E_3>0$ with $0<\Omega^{{\text{\rm{Is}}}} <\pi/4$.}\label{fig1}

\end{figure}

Very recently it has indeed been proven \cite{dohm-wessel-2021} that two-scale-factor universality is violated for the weakly anisotropic scalar $\varphi^4$ model with periodic boundary conditions (PBC) in $d=2$ dimensions. An exact result has been derived for ${\cal F}_c(q,\Omega)$ of the general $\varphi^4$ model with arbitrary short-range pair interactions, i.e., for arbitrary orientations $\Omega$ of the principal axes, by means of conformal field theory (CFT) where no assumption has been made other than the validity of two-scale-factor universality for ${\it isotropic}$ systems. Furthermore, on the basis of the hypothesis of multiparameter universality, quantitative predictions have been made for ${\cal F}^{{\text{\rm{Is}}}}_c={\cal F}_c(q^{{\text{\rm{Is}}}},\Omega^{{\text{\rm{Is}}}})$ of the fully anisotropic triangular Ising model with PBC for the case of arbitrary orientations $\Omega^{{\text{\rm{Is}}}}$ of the principal axes  \cite{dohm-wessel-2021}.
The Hamiltonian of this model reads
\cite{dohm2019,Indekeu,stephenson}
\begin{equation}
\label{IsingH}
H^{{\text{\rm{Is}}}}\! =\!-\!\!\sum_{j, k}\big [E_1\sigma_{j,k} \sigma_{j,k+1}+E_2\sigma_{j,k} \sigma_{j+1,k}  +E_3\sigma_{j,k} \sigma_{j+1,k+1}\big]
\end{equation}
where $\sigma_{j,k}=\pm1$ are spin variables on the square lattice
(with the lattice spacing $\tilde a = 1$)
with horizontal, vertical, and diagonal couplings $E_1, E_2$, $E_3$ (see Fig.~\ref{fig1}). Both the  angle
$\Omega^{\rm{Is}}(E_1,E_2,E_3)$ of the principal axes and the  ratio of the principal correlation lengths $q^{\rm{Is}}(E_1,E_2,E_3)=\xi_{0\pm}^{(1)\rm{Is}}/\xi_{0\pm}^{(2)\rm{Is}}$  are known
functions of
$E_1,E_2,E_3$ \cite{dohm2019}. The angle is given by \cite{dohm2019}
\begin{eqnarray}
\label{omega1}
\tan 2\Omega^{\rm{Is}}&=&\frac{2\hat S_3}{\hat S_1- \hat S_2}\;\; \; \mbox{for} \;\; E_1\neq E_2,\;\;\;\;\;\;\;
\\
\label{omega2}
\Omega^{\rm{Is}}&=&\pi/4\;\;\;\;\; \;\;\;\;\;  \mbox{for} \;\; \;E_1=E_2,
\end{eqnarray}
with
%
%
$\hat S_\alpha= \sinh 2\beta^{\rm{Is}}_c E_\alpha, \alpha=1,2,3, \beta^{\rm{Is}}_c= (k_B T^{\rm{Is}}_c)^{-1}$.

Exact agreement was reported between the prediction of \cite{dohm-wessel-2021} and the exact results \cite{izmail,HH2019} for ${\cal F}^{{\text{\rm{Is}}}}_c$ of the anisotropic Ising model with a "rectangular anisotropy" $E_1\neq E_2$, $E_3=0$ where the principal axes  are parallel to the symmetry axes of the lattice, i.e., $\Omega^{{\text{\rm{Is}}}}=0$ \cite{dohm2019}, thus confirming multiparameter universality for this case. However, so far no proof exists for the validity of multiparameter universality for finite anisotropic Ising models with a nonzero angle $\Omega^{{\text{\rm{Is}}}}$ of the principal axes.
\vspace{0cm}

It is the purpose of this paper to test the predictions of \cite{dohm-wessel-2021} for ${\cal F}^{{\text{\rm{Is}}}}_c$ for the case of a square geometry with a finite angle $\Omega^{{\text{\rm{Is}}}}=\pi/4$, i.e where the principal axes are parallel to the diagonals of the square lattice (Fig. 1). This will be achieved by means of high-precision Monte Carlo (MC) simulations for ${\cal F}^{{\text{\rm{Is}}}}_c$ of the Ising model with the Hamiltonian (\ref{IsingH}) on a square lattice with isotropic ferromagnetic couplings $E_1=E_2=E>0$ between nearest neighbors (NN) and with an anisotropic coupling $E_3\neq0$ between next-nearest neighbors (NNN) along one diagonal in the ferromagnetic region $E+E_3>0$ \cite{Berker}. We find remarkable agreement between the MC data and the analytical prediction. The comparison with the corresponding result for the scalar $\varphi^4$ model supports the validity of multiparameter universality but invalidates two-scale-factor universality as ${\cal F}^{{\text{\rm{Is}}}}_c$ is found to exhibit a nonuniversal dependence on the ratio $E_3/E$. That differs from the corresponding dependence of ${\cal F}_c$ of the $\varphi^4$ model although the latter model and the Ising model belong to the same
universality class.

\renewcommand{\thesection}{\Roman{section}}
\renewcommand{\theequation}{2.\arabic{equation}}
\setcounter{equation}{0}
\section{ Shear transformation of the anisotropic $\varphi^4$ model}
We first consider the anisotropic scalar $\varphi^4$ lattice model with dimensionless continuous variables $-\infty \leq \varphi_i\leq \infty$ on $N=(L/\tilde a)^2$ lattice points ${\bf x}_i \equiv (x_{i1}, x_{i2})$ on a square lattice with lattice spacing $\tilde a$ and a shape of an $L \times L$ square with PBC. We assume short-range pair interactions $K_{i,j}$. The Hamiltonian and the free-energy density divided by $k_B T$ are defined by \cite{dohm2006,dohm2008}
\begin{eqnarray}
\label{2a}&& H  =   \tilde a^2 \Bigg[\sum_{i=1}^N \left(\frac{r_0}{2}
\varphi_i^2 + u_0 \varphi_i^4 \right)
 +\sum_{i, j=1}^N \frac{K_{i,j}} {2} (\varphi_i -
\varphi_j)^2 \Bigg], \;\;\;\;\\
\label{free energy}
 &&f(t, L) =  - L^{-2} \ln \Bigg\{ \Big[\prod_{i = 1}^N \int_{-\infty}^\infty
d \varphi_i \Big] \exp \left(- H \right)\Bigg\}.
\end{eqnarray}
The large-distance anisotropy is described by the $2 \times 2$ symmetric anisotropy matrix ${\bf A}$ with matrix elements determined by the  second moments
\begin{eqnarray}
\label{2i} {\bf A}_{\alpha\beta} &=&\lim_{N \to \infty} N^{-1} \sum_{i,
j} (x_{i \alpha} - x_{j \alpha}) (x_{i \beta} - x_{j \beta})
K_{i,j} .\;\;\;\;
\end{eqnarray}
We assume the following anisotropy matrix
\begin{eqnarray}
\label{abc}
{\bf A}&=&({\bf A}_{\alpha\beta} )
=
\left(\begin{array}{ccc}
 a & c \\
 c & a \\
\end{array}\right)
\end{eqnarray}
%
with $c\neq 0$ and positive diagonal elements ${\bf A}_{11}={\bf A}_{22}=a>0$.
Weak anisotropy requires  positive eigenvalues $\lambda_1=a+c>0,\lambda_2=a-c>0$, i.e. $-a< c <a$, which
ensures unchanged critical exponents of the two-dimensional Ising universality class \cite{cd2004}. The eigenvectors
%
\begin{eqnarray}
 \label{2pp}
 {\bf e}^{(1)}& =& \frac{1}{\sqrt{2}}\left(\begin{array}{c}
  1 \\
  1 \\
\end{array}\right) ,\;\;\; {\bf e}^{(2)} =\frac{1}{\sqrt{2}} \left(\begin{array}{c}
  - 1 \\
  1 \\
\end{array}\right) \;\;\;
\end{eqnarray}
are valid for  both $c>0$ and $c<0$. The principal directions 1 and 2 are parallel and perpendicular to the $(1,1)$ direction corresponding to the angle $\Omega=\pi/4$. Near $T_c$ the asymptotic principal correlation lengths
$\xi_\pm^{(\beta)}(t)= \xi^{(\beta)}_{0\pm}|t|^{-1}$
have the ratio \cite{dohm2008,dohm2018}
\be
\label{ratioq} \frac{\xi_{0 \pm}^{(1)}}{\xi_{0 \pm}^{(2)}} = \Big(\frac{\lambda_1}{\lambda_2}\Big)^{1/2}=\Big(\frac{a+c}{a-c}\Big)^{1/2}
=  q   = \left\{
\begin{array}{l}
               > 1, \quad c > 0, \\
               < 1 ,\quad  c < 0.
                \end{array} \right.
\ee
The simplest version of this $\varphi^4$ model is realized \cite{dohm2006} by isotropic ferromagnetic NN couplings  $K =J/\tilde a^2>0$ in the horizontal and vertical directions and a finite anisotropic NNN coupling $K_d=J_d/\tilde a^2$ in the diagonal $(1,1)$ direction [Fig. 2 (a)]. This yields the dimensionless nondiagonal anisotropy matrix \cite{dohm2006,dohm2018}
\begin{eqnarray}
\label{matrixA}
{\bf A}
=
2\left(\begin{array}{ccc}
J +J_d& J_d \\
 J_d & J +J_d \\
\end{array}\right).
\end{eqnarray}
with  $0 < {\det \bf  A}< \infty$ and
\begin{eqnarray}
\label{qratio}
q=(1+2J_d/J)^{1/2}= \left\{
\begin{array}{l}
               > 1, \quad J_d > 0, \\
               <1 , \quad J_d < 0,
                \end{array} \right.
\end{eqnarray}

in the range $-J/2<J_d< \infty$. We are interested in the asymptotic (large $L$) amplitude ${\cal F}_c= L^2f_{sc}$ where $f_{sc}\equiv f_s(0,L)$ is the singular part of $f(t,L)$ at $T_c$. In the following we specialize the general derivation of \cite{dohm-wessel-2021} to the present case.

From the structure of previous results for the anisotropic $\varphi^4$ theory in $2\leq d <4$ dimensions {\cite{cd2004,dohm2006,dohm2008,kastening-dohm,dohm2018,dohm2019,chen-zhang} it is known that observable anisotropy effects are described by the reduced anisotropy matrix  ${\bf \bar A}= {\bf A}/(\det {\bf A})^{1/d}$ which in the present two-dimensional case has the form
\begin{eqnarray}
  \label{Abarx}
 {\bf \bar A}(q)=
 \;\frac{1}{2} \left(\begin{array}{ccc}
    q+q^{-1} &\;\;\; q-q^{-1}  \\
  q-q^{-1} & \;\;\;q+q^{-1}  \\
\end{array}\right).\;\;
\end{eqnarray}
This matrix has the same  eigenvectors (\ref{2pp}) as ${\bf A}$ and the reduced eigenvalues $\bar \lambda_\beta =\lambda_\beta/(\det {\bf A})^{1/2}$, $\beta=1,2,$ with  $\det {\bf \bar A}=\bar \lambda_1 \bar \lambda_2 =1$. It is expected that ${\cal F}_c$ is a function of ${\bf \bar A}(q)$,
\begin{equation}
\label{critampq}
{\cal F}_c(q)={\cal F}_c[{\bf \bar A}(q)].
\end{equation}
It has been shown \cite{cd2004,dohm2006,dohm2008} that a shear transformation can be performed such that the anisotropic $\varphi^4$ model on a square is transformed to a $\varphi^4$ model on a rhombus (Fig.~\ref{fig:shear}) with changed second moments ${\bf A'}_{\alpha \beta}= \delta_{\alpha \beta}, {\bf A'}={\bf 1}$, representing a system with isotropic critical correlations at large distances.
The couplings $K_{i,j}$ , $K$, and $K_d$ and the temperature variable $r_0$ are kept fixed in this transformation which is constructed such that the Hamiltonian is invariant.
We describe the vertical sides of the square and the corresponding transformed sides of the rhombus by the vectors $\bm L$ and $\bm{L}_\text{rh}$, respectively (Fig. ~\ref{fig:shear}). They are related by the transformation
\begin{eqnarray}
&&\bm{L}_\text{rh}={\bf {\mbox {\boldmath$\lambda$}}}^{-1/2}\bm{U}{\bm L}= \frac{L}{2^{1/2}}\left(\begin{array}{c}
  \lambda_1^{-1/2} \\
  \lambda_2^{-1/2}  \\
\end{array}\right),{\bm L}=L \left(\begin{array}{c}
  0 \\
  1  \\
\end{array}\right),\;\;\;\;\;\;\\
%
%
&&{\bf U}=\frac{1}{2^{1/2}}\left(\begin{array}{ccc}
  1& 1 \\
  -1 &  1 \\
\end{array}\right),
{\bf {\mbox {\boldmath$\lambda$}}}={\bf U AU}^{-1}
=\left(\begin{array}{ccc}
 \lambda_1& 0 \\
 0 & \;\lambda_2 \\
\end{array}\right),\;\;\;\;\;\;\;
\end{eqnarray}
with the rotation matrix ${\bf U}$  and the diagonal rescaling matrix ${\bm \lambda}$.
Here we have considered a clockwise rotation; equivalent results are obtained by a counterclockwise rotation.
The ratio of the components of $\bm{L}_\text{rh}$ determines the angle $0< \omega < \pi$ (Fig.~\ref{fig:shear} b) as $(\lambda_2/\lambda_1)^{1/2}= \tan(\omega/2)$ or \cite{dohm2006,dohm2008}
\begin{eqnarray}
\label{angle}
 \omega(q)&=&2 \arctan\Big[\Big(\lambda_2/\lambda_1\Big)^{1/2}\Big]\\
 \label{anglecorr}
 &=&2 \arctan\xi_{0 \pm}^{(2)}/\xi_{0 \pm}^{(1)}\;\;
=  2 \arctan q^{-1}.
\end{eqnarray}
Instead of $\omega$ we may employ the complementary angle $0 < \alpha=\pi - \omega < \pi$ which yields
\begin{eqnarray}
 \label{alpha}
\alpha(q) &=&   2 \arctan q.
\end{eqnarray}
The singular part ${\cal F}^{\rm{iso}}_{c\rm{}}(\alpha)$ of the free energy at $T_c$ of the rhombus is a function of $\alpha$. The total free energies of the systems on the square and on the rhombus differ only by a nonsingular additive term \cite{dohm2006}. Thus the singular part ${\cal F}_c$ of the anisotropic system is left invariant and is determined by
\begin{eqnarray}
\label{freeEnergyanisox}
{\cal F}_c(q)
={\cal F}^{\rm{iso}}_{c}\big(\alpha(q)\big).
\end{eqnarray}
%
However, it is unknown how to perform an exact calculation of ${\cal F}^{\rm{iso}}_{c}(\alpha)$ within the isotropic $\varphi^4$ model on a rhombus.

\begin{figure}[t!]
\begin{center}
\includegraphics[width=\columnwidth]{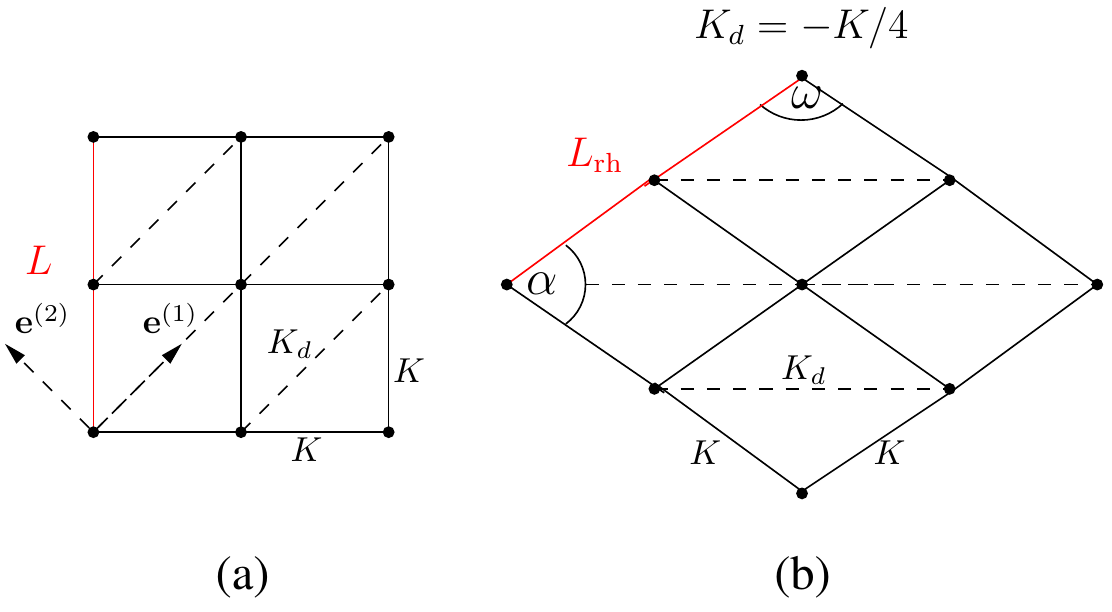}
\end{center}
\caption{
(a) Square lattice with isotropic NN couplings $K$ and anisotropic NNN coupling $K_d$. (b) For the case $K_d=-K/4$ corresponding to $\lambda_1/\lambda_2=1/2$, the transformed lattice with isotropic second moments $A'_{\alpha\beta}=\delta_{\alpha\beta}$ has the angle $\omega = 2\arctan(\sqrt{2})$ and the complementary angle $\alpha=2\arctan(1/\sqrt{2})$ (compare Fig. 2 of \cite{dohm2006}).
}
\label{fig:shear}
\end{figure}
%
\renewcommand{\thesection}{\Roman{section}}
\renewcommand{\theequation}{3.\arabic{equation}}
\setcounter{equation}{0}
\section{ Exact result for the anisotropic $\varphi^4$ model}
At this point we follow the reasoning of \cite{dohm-wessel-2021} where two-scale-factor universality for isotropic systems \cite{pri} is invoked. This means that isotropic scalar $\varphi^4$ models and isotropic Ising models have the same universal amplitude ${\cal F}^{\rm{iso}}_c={\cal F}^{\rm{Is},iso}_{c\rm{}}$ if they are defined on a rhombus with the same angle $\alpha$ and with the same BC. From conformal field theory (CFT) \cite{franc1987,franc1997} an exact asymptotic contribution $Z^{\rm{CFT}}(\tau)$ to the partition function of the isotropic Ising model at $T_c$ on a rhombus with PBC, i.e., on a torus, has been derived where the rhombus is parameterized by a complex torus modular parameter
\begin{eqnarray}
\label{tau}
\tau(\alpha) = \mbox{Re}\; \tau + i\; \mbox{Im} \;\tau=\exp(i\; \alpha)
\end{eqnarray}
with $|\tau|=1,\mbox{Im} \;\tau> 0$,   and with the angle $0< \alpha < \pi$ (see Fig.~\ref{fig:cft}),
\be
\label{anglealpha}
\alpha= \arctan(\mbox{Im}\; \tau/\mbox{Re}\; \tau).
\ee
%
\begin{figure}[t!]
\begin{center}
\includegraphics[width=\columnwidth]{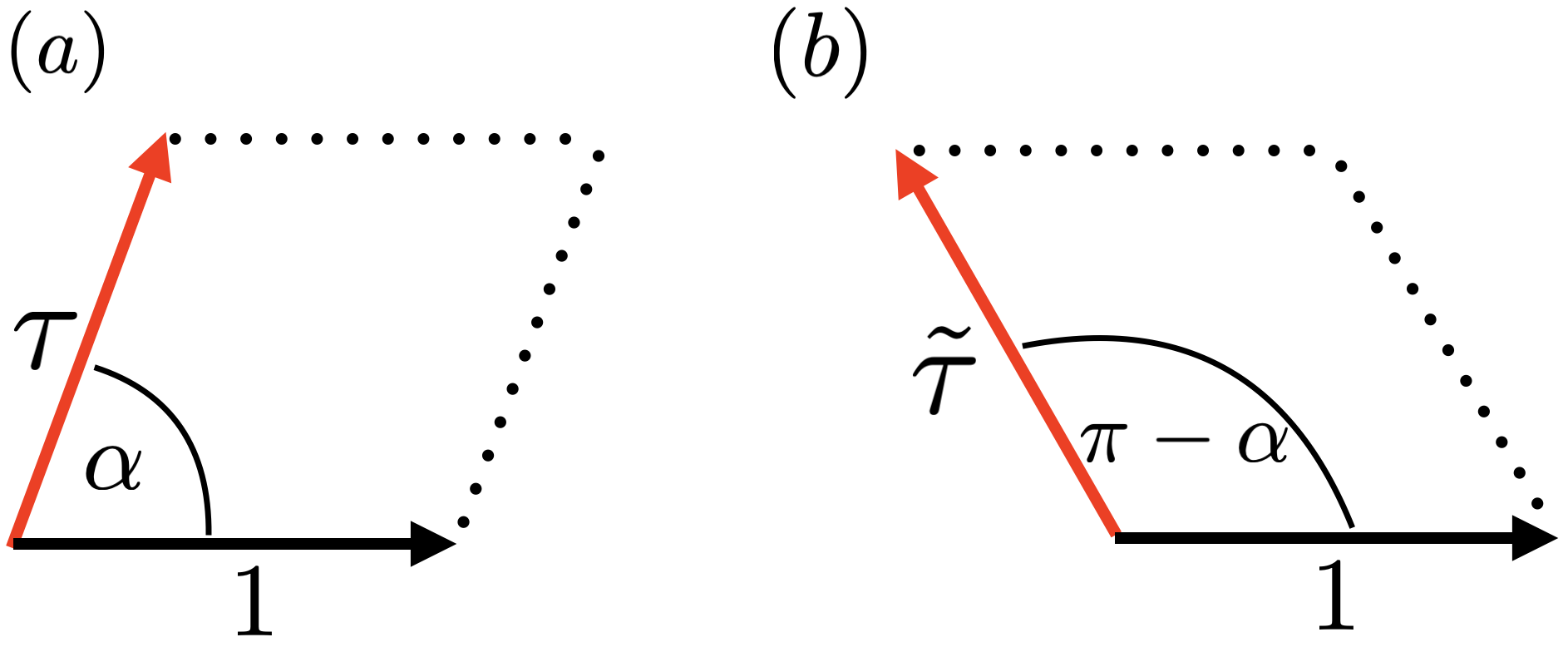}
\end{center}
\caption{
Illustration of the two choices for mapping the anisotropic square onto an isotropic rhombus with a clockwise rotation (compare Fig.~\ref{fig:shear}) or a counterclockwise rotation.
Indicated in panel (a) are the angle $\alpha=2\arctan(1/\sqrt{2})$
and  the modular parameter $\tau$, and in panel (b) the complementary angle $\pi-\alpha=2\arctan(\sqrt{2})$ and  $\tilde\tau=-1/\tau$ .
}
\label{fig:cft}
\end{figure}
%
The partition function is expressed in terms of Jacobi theta functions $\theta_i(0|\tau)\equiv \theta_i(\tau)$ (in the notation of \cite{franc1997}, see App.~\ref{appa}) as~\cite{franc1987}
\begin{equation}
\label{ZIsing}
Z^{\rm{CFT}}(\tau)=\big({|\theta_2(\tau)|+|\theta_3(\tau)|+|\theta_4(\tau)|}\big)/\big({2|\eta(\tau)|}\big),
\end{equation}
with
$\eta(\tau)=(\frac{1}{2}\theta_2(\tau)\theta_3(\tau)\theta_4(\tau))^{1/3}$, from which we obtain
\begin{equation}
\label{FCFT}
{\cal F}^{\rm{CFT}}_{c}(\tau)=\!-\ln Z^{\rm{CFT}}(\tau).
\end{equation}
The singular part of the free energy of the isotropic Ising model at $T_c$ is
\begin{eqnarray}
\label{freeCFT}
{\cal F}^{\rm{Is},iso}_{c\rm{}}\!\big(\tau(\alpha)\big)
\!={\cal F}^{\rm{CFT}}_c\!\big(\tau(\alpha)\big)
\!= {\cal F}^{\rm{iso}}_{c\rm{}}\!\big(\alpha\big),
\end{eqnarray}
where, owing to two-scale-factor universality, the last equation applies to the transformed $\varphi^4$ model on the rhombus. We define
\begin{eqnarray}
\label{tauq}
 \tau[q]= \tau\big(\alpha(q)\big)=\exp\big(i\; \alpha(q)\big)\nonumber\\=  \frac{q^2-1}{q^2+1}+  i \frac{2q}{1+q^2}.
\end{eqnarray}
Then we obtain from (\ref{freeCFT}) and (\ref{freeEnergyanisox}) our exact result for the critical amplitude ${\cal F}_c$ of the anisotropic $\varphi^4$ model as
\begin{eqnarray}
\label{calFCFT}
{\cal F}_c(q)={\cal F}_c^{\rm{CFT}}\big(\tau[q]\big)
\end{eqnarray}
%
where the  nonuniversal expressions (\ref{ratioq}) or (\ref{qratio}) for $q(c/a)$  or $q(J_d/J)$ have to be inserted. So far our only assumption is the validity of two-scale-factor universality for isotropic systems.

We note that a counterclockwise rotation in the shear transformation replaces $\tau$ by
$\widetilde{\tau}=-1/\tau=-\tau^*=-\mbox{Re}\;\tau+i\;  \mbox{Im}\;\tau$ (Fig.~\ref{fig:cft}), using that $|\tau|=1$. The modular invariance $Z^{\rm{CFT}}(\tau)=Z^{\rm{CFT}}(-1/\tau)$~\cite{franc1987,franc1997}  guarantees  the independence of ${\cal F}^{\rm{CFT}}_c\big(\tau[q]\big)$ on this choice.
This invariance also implies that $Z^{\rm{CFT}}$ only depends on  $|\mbox{Re}\; \tau|$ under the condition $|\tau|=1$, $1/\tau=\tau^*$. Thus,
the critical amplitude  ${\cal F}^{\rm{CFT}}_c\big(\tau[q]\big)$  is a function of $s(q)^2$, i.e., a symmetric function of
\begin{eqnarray}
\label{sq1}
s(q)&&\equiv \mbox{Re}\; \tau[q]= (q^2-1)/(q^2+1)\\
\label{sq2}
&&= (1+J/J_d)^{-1}= \left\{
\begin{array}{l}
               > 0, \quad J_d > 0, \\
               <0 , \quad J_d < 0,
                \end{array} \right.
\end{eqnarray}
%
which is shown
in Fig.~\ref{fig:Fc}.
Since $q$ depends on the microscopic couplings the amplitude ${\cal F}_c(q)$ is a nonuniversal quantity violating two-scale-factor universality.
%
\begin{figure}[t!]
\begin{center}
\includegraphics[width=\columnwidth]{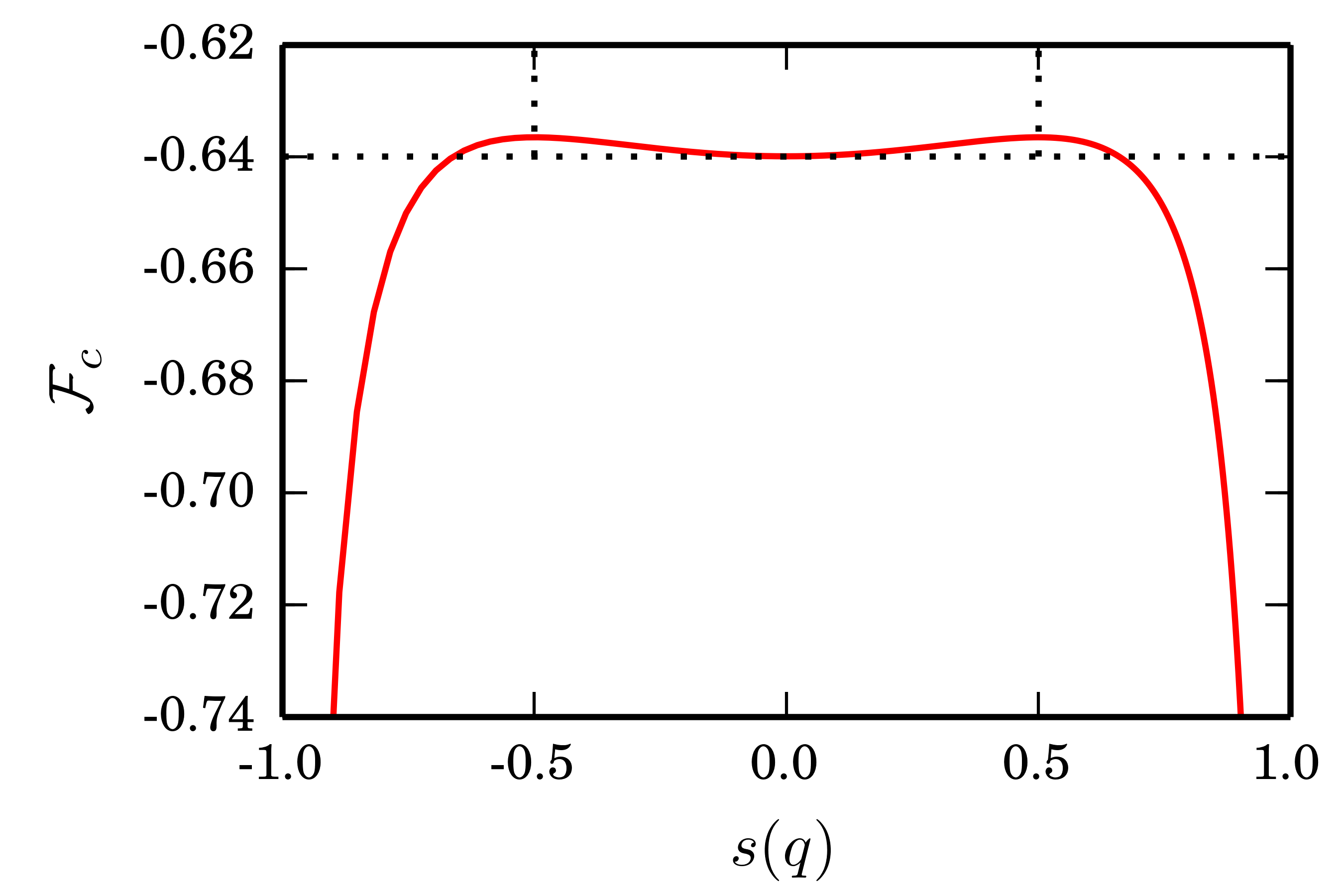}
\end{center}
\caption{
Exact analytic result ${\cal F}_c(q)$, (\ref{calFCFT}) for the anisotropic $\varphi^4$ theory at $T=T_c$ as a symmetric function  of $ s(q)$, (\ref{sq1}) with maxima at $s(q)=\pm 1/2$, indicated by the vertical lines.
The horizontal line represents the isotropic value of ${\cal F}_c(1)=-0.639912.$ for $q=1$ ($s(q)=0, J_d=0$).  Positive (negative) values $s(q)>0$ ($s(q)<0$) correspond to the range of ferromagnetic (antiferromagnetic) couplings $J_d$.  A symmetric representation is also possible as a function of $\ln q$.}
\label{fig:Fc}
\end{figure}
%
Since positive (negative) values of $s(q)$ correspond to $J_d>0$ ($J_d<0$) according to (\ref{sq2}) we see from Fig.~\ref{fig:Fc} that ${\cal F}_c$ yields the same values in the range of ferromagnetic and antiferromagnetic couplings $J_d$. This symmetry of ${\cal F}_c$  only reflects the modular invariance in the presence of PBC but does not yet imply the  universality of the functional form of ${\cal F}_c(q)$.  In our present case, the function ${\cal F}_c$ exhibits a
two-peak structure
where the two equal-height maxima  are located at
$s(q_{\mathrm{max}})=\pm\frac{1}{2}$
%
corresponding to $q_{\mathrm{max}}=3^{\pm1/2}$ and $J_d=J, J_d=-J/3$, respectively.
Furthermore, we note that  ${\cal F}_c\rightarrow - \infty$ for $q\rightarrow 0$ ($s(q)\rightarrow -1$)  and  $q\rightarrow \infty$ ($s(q)\rightarrow 1$), where
the modular parameter approaches $\tau\rightarrow -1$ and  $\tau\rightarrow +1$ respectively, in these limits.
In terms of  coupling parameters, these limits correspond to $J_d \rightarrow -J/2$ and
$J_d\rightarrow \infty$, respectively.
Hence, the CFT rhombus degrades towards both ends of the regime $-1/2<J_d/J< \infty$  of weak anisotropy. By rewriting $q=\exp(\ln q)$ we note that one finds that ${\cal F}_c(q)=\widetilde {\cal F}_c(\ln q)$ is a symmetric function also of $\ln q$, $\widetilde {\cal F}_c(\ln q)=\widetilde {\cal F}_c(-\ln q)$, as a consequence of modular invariance.

%
\renewcommand{\thesection}{\Roman{section}}
\renewcommand{\theequation}{4.\arabic{equation}}
\setcounter{equation}{0}
\section{ Analytic prediction and Monte Carlo simulation for the anisotropic Ising model}
We return to the Ising model defined by (\ref{IsingH}) in a square geometry.
 The free energy per site (divided by $k_B T$) for a square lattice with $L^2$ lattice sites at $\beta=1/(k_B T)$  is
\begin{equation}
\label{freeIsing}
f^\mathrm{Is}{(\beta,L)}= -\frac{1}{L^{2}}\ln \sum_{\{\sigma\}}\exp(- \beta H^\mathrm{Is}).
\end{equation}
The hypothesis of multiparameter universality \cite{dohm2018} for the confined system predicts that, parallel to the case for the bulk correlation function \cite{dohm2019}, the singular part ${\cal F}^\mathrm{Is}_c$ of the total free energy $L^2 f^\mathrm{Is}{(\beta^{\rm{Is}}_c,L)}$ at $T^\mathrm{Is}_c$  is obtained from ${\cal F}_c(q,\Omega)$ of the scalar $\varphi^4$ model at $T_c$ by the substitution $ q \to q^\mathrm{Is}, \Omega \to \Omega^\mathrm{Is}$  provided that both models have the same geometry and the same BC. Thus the prediction is \cite{dohm-wessel-2021}
%
${\cal F}^{\rm{Is}}_c =  {\cal F}_c(q^{\rm{Is}},\Omega^{\rm{Is}})$.
%
Here we consider this Ising model for $E_1=E_2=E>0, E_3\neq 0$ in the ferromagnetic range $E+E_3>0$ \cite{Berker} with $\Omega^\mathrm{Is}=\pi/4$ and the exact ratio of the principal correlation lengths \cite{dohm2019}
\be
\label{ratioqising}
q^{\rm{Is}}=\frac{\xi_{0 \pm}^{(1)\rm{Is}}}{\xi_{0 \pm}^{(2)\rm{Is}}} = \frac{1}{\sinh 2\beta^{\rm{Is}}_c E} = \left\{
\begin{array}{l}
               > 1, \quad E_3 > 0, \\
               <1 , \quad E_3 < 0.
                \end{array} \right.
\ee
Due to the condition of criticality \cite{Berker}
\be
\label{critising}
\sinh^2(2\beta^{\rm{Is}}_c E)+ 2\sinh(2\beta^{\rm{Is}}_c E)\sinh(2\beta^{\rm{Is}}_c E_3)=1,
\ee
$2\beta^{\rm{Is}}_c E\equiv y$ is a function of $E_3/E$ determined implicitly by
%
$\sinh^2(y)+ 2\sinh(y)\sinh(yE_3/E)=1$,
thus $q^{\rm{Is}}=1/\sinh(y)$ is a function of $E_3/E$ in the range $-1<E_3/E<\infty$.
In the present case ($\Omega^\mathrm{Is}=\pi/4$), Eq. (\ref{critampq}) is replaced by
\begin{equation}
\label{critampqIs}
{\cal F}^{\rm{Is}}_c(q^{\rm{Is}})={\cal F}_c[{\bf \bar A}(q^{\rm{Is}})]
\end{equation}
where ${\bf \bar A}$ is the same matrix as in (\ref{Abarx}), with $q$ replaced by $q^{\rm{Is}}$. Thus our prediction for the Ising model is
\begin{eqnarray}
\label{calFCFTIs}
{\cal F}^{\rm{Is}}_c(q^{\rm{Is}})={\cal F}_c^{\rm{CFT}}\big(\tau[q^{\rm{Is}}]\big)
\end{eqnarray}
where ${\cal F}_c^{\rm{CFT}}$ is the same function as in (\ref{calFCFT}) with $q$ replaced by $q^{\rm{Is}}$.
%
\begin{figure}[t!]
\begin{center}
\includegraphics[width=\columnwidth]{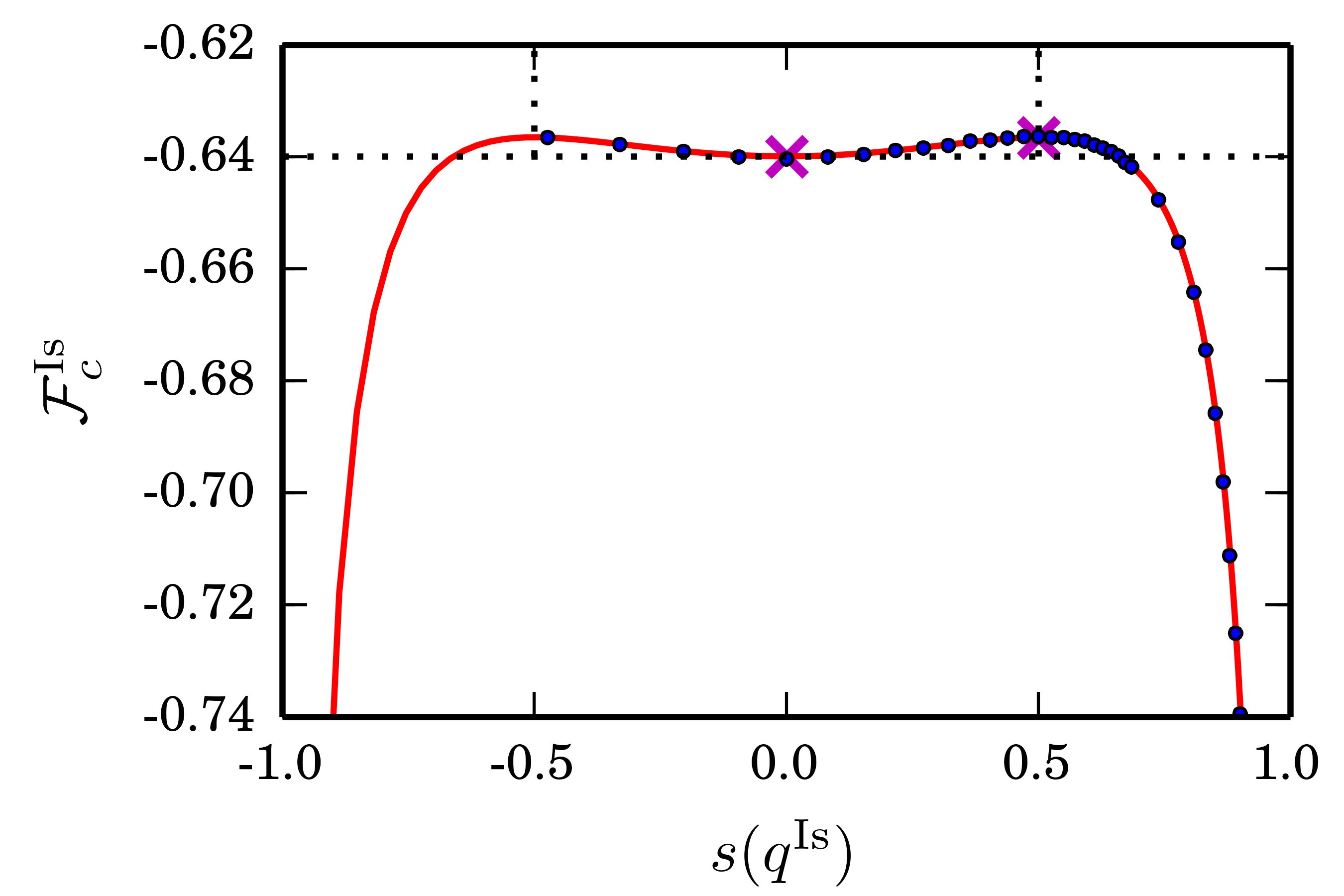}
\end{center}
\caption{
Comparison of the
exact analytic prediction ${\cal F}_c^\text{Is}$, (\ref{calFCFTIs}) for the anisotropic Ising  model,(\ref{IsingH}), to
finite-size estimates ${\cal F}^\text{Is}_{c,L}$ from MC simulations for $L=16$ (circles) as symmetric functions
of $ s(q^\text{Is})$, with maxima at $s(q^\text{Is})=\pm 1/2$, indicated by the vertical lines.
The statistical uncertainty of the MC data is below the  symbol size.
The crosses at $s(q^\text{Is})=0$ and $s(q^\text{Is})=1/2$ are the analytical results from Refs.~\cite{FF} and  \cite{salas}, respectively.
The horizontal line represents the isotropic value of ${\cal F}^\text{Is}_c(1)=-0.639912$  for $q^\text{Is}=1$ ($s(q^\text{Is})=0, E_3=0$). The positive (negative) values $s(q^\text{Is})>0$ ($s(q^\text{Is})<0$) correspond to the range of ferromagnetic (antiferromagnetic) couplings $E_3$. A symmetric representation is also possible as a function of $\ln q^\text{Is}$.}
\label{fig:FcIs}
\end{figure}
%
${\cal F}^{\rm{Is}}_c$ is plotted in Fig.~\ref{fig:FcIs} as a function of
\begin{eqnarray}
\label{sq1Ising}
s(q^{\rm{Is}})&&\equiv \mbox{Re}\; \tau[q^{\rm{Is}}]= [(q^{\rm{Is}})^2-1]/[(q^{\rm{Is}})^2+1]\\
\label{sq2Ising}
&&= \frac{1-\sinh^2(2\beta^{\rm{Is}}_c E)}{\cosh^2(2\beta^{\rm{Is}}_c E)}= \left\{
\begin{array}{l}
               > 0, \quad E_3 > 0, \\
               <0 , \quad E_3 < 0,
                \end{array} \right.
\end{eqnarray}
in terms of which it exhibits the exact identical form as does ${\cal F}_c$ as a function of $s(q)$ in Fig.~\ref{fig:Fc}.
%
%
%
A significant analytic test of the prediction  (\ref{calFCFTIs}) is
provided by a comparison with available exact results for isotropic
Ising models on a square lattice with the shape of a square \cite{FF}
and on a triangular lattice with three equal NN couplings with the shape
of a rhombus \cite{salas} (crosses in Fig.~\ref{fig:FcIs}).
The former model is identical with our model with $E_3=0$, the latter
isotropic model is topologically equivalent to our anisotropic model (with the shape
of square) with $E_3=E$.
Indeed, our formula for  ${\cal F}^\text{Is}_c(q^\text{Is})$ at $s(q^\text{Is})=0$ ($q^\text{Is}=1$) and $s(q^\text{Is})=1/2$ ($q^\text{Is}=\sqrt{3}$) exactly reduces  to the  analytic results stated there, given that these lattices correspond to $\tau=i$ and $\tau=e^{i\pi/3}$, respectively, as noted in \cite{dohm-wessel-2021}.
We also note that  ${\cal F}^\text{Is}_c\rightarrow - \infty$ for $E_3\rightarrow -E$ and $E_3\rightarrow \infty$ (corresponding to $q^\text{Is}\rightarrow 0$ ($s^\text{Is}\rightarrow -1$ ) and  $q^\text{Is}\rightarrow \infty$ ($s^\text{Is}\rightarrow +1$)) , as the modular parameter approaches $\tau\rightarrow -1$ and  $\tau\rightarrow +1$. Thus, for the Ising model, the CFT rhombus degrades towards both ends of the regime $-1<E_3/E< \infty$ of finite $T_c$. While in the upper limit $E_3/E \to \infty$, the system decouples into one-dimensional chains, it does not  exhibit order even at $T=0$, due to  frustration  for $E_3/E<-1$  beyond the lower limit.

In order to assess the validity of our analytical prediction  for  general values of $E_3/E$, we performed  MC simulations of the Hamiltonian $H^\text{Is}$ on finite $L\times L$ lattices with periodic BC,
using local spin flips with the standard Metropolis update,
combined (for $E_3\geq 0$) with Wolff-cluster updates~\cite{wolff}.
We obtain the free energy per site at $T^{\rm{Is}}_c$, divided by $k_B T^{\rm{Is}}_c$ , $f^\text{Is}{(\beta^{\rm{Is}}_c,L)}$, from a thermodynamic integration
\begin{equation}
f^\text{Is}{(\beta^{\rm{Is}}_c,L)}=\int_0^{\beta^\text{Is}_c} u^\text{Is}(\beta,L)\: d\beta - \ln 2
\end{equation}
of the inner energy per site,  $u^\text{Is}(\beta,L)$, over a dense inverse
temperature grid,
using the trapezoidal rule to numerically perform the above integral.
We
used a uniform $\beta$-grid with $\Delta\beta = 0.0005/E$ for all reported data. The above integration formula is obtained from
$u^\text{Is}{(\beta,L)} = \partial f^\text{Is}{(\beta,L)}/\partial\beta $ and $f^\text{Is}(0,L)=-\ln 2$,
as follows
from (\ref{freeIsing}).
As a finite-size estimate for ${\cal F}^\text{Is}_c$ we  consider the quantity
\begin{equation}
{\cal F}^\text{Is}_{c,L}=\frac{4}{3}{[f^\text{Is}(\beta^{\rm{Is}}_c,L)-f^\text{Is}(\beta^{\rm{Is}}_c,2L)]} L^2
\end{equation}
which, by the asymptotic scaling (\ref{freeamplitude}), approaches ${\cal F}^\text{Is}_c = \lim_{L\to \infty}{\cal F}^\text{Is}_{c,L}$ in the thermodynamic limit. A comparison of the MC data to our analytical prediction of ${\cal F}^\text{Is}_c$ is shown in Fig.~\ref{fig:FcIs}
as a function of  $s(q^\text{Is})$.
Here we show data for $L=16$ where we were able to obtain accurate values for ${\cal F}^\text{Is}_{c,L}$. The residual finite-size effects are found to be weak, as seen from comparing to the exact previous values at $E_3=0$ and $E_3=E$~\cite{FF,salas}.
Aside from  weak residual finite-size effects, we obtain a remarkable  agreement between the MC data and the analytical prediction over the full parameter range  for which we could obtain  high-precision MC data.
The statistical uncertainty from the MC simulations significantly grows for $E_3\lesssim -0.5E$, corresponding to $s(q^\text{Is})\lesssim -0.5$ due to the enhanced competition between the antiferromagnetic coupling $E_3<0$ and the ferromagnetic couplings $E_1=E_2=E>0$.

The agreement between the MC data and the analytical prediction of ${\cal F}^\text{Is}_c$ in the accessible range $E_3\gtrsim -0.5E$ provides substantial support for the validity of multiparameter universality for the case $\Omega^{\rm{Is}}=\pi/4$. The symmetry of ${\cal F}^\text{Is}_c(q^\text{Is})$ with respect to the ferromagnetic ($E_3>0)$ and antiferromagnetic ($E_3<0)$ range of the coupling $E_3$ predicted by the theory is well reproduced by the MC data. In particular, two equal-height maxima  at $s(q_{\text{max}}^\text{Is})=\pm{1}/{2}$ are detected corresponding to the couplings $E_3=E$ and $E_3=-[\text{arcsinh}(1/\sqrt3)/ \text{arcsinh}(\sqrt3)]E\approx -0.4171E $. We note that this symmetric structure results from the modular invariance discussed in the context Fig. 4 and is not expected to be a generic feature of multiparameter universality for the general case of a rectangular geometry and $\Omega^{\rm{Is}}\neq\pi/4$ ~\cite{dohm-wessel-2021}.
\renewcommand{\thesection}{\Roman{section}}
\renewcommand{\theequation}{5.\arabic{equation}}
\setcounter{equation}{0}
\section{ Violation of two-scale-factor universality}
Due to multiparameter universality, the representation of ${\cal F}^{\rm{Is}}_c$ as a function of the ratios $q$ and $q^{\rm{Is}}$ of principal correlation lengths is predicted  \cite{dohm-wessel-2021,dohm2018} to have a universal character implying that ${\cal F}^{\rm{Is}}_c(q^{\rm{Is}})={\cal F}_c(q^{\rm{Is}})$, i.e.,  with the same functional form of ${\cal F}_c$ for the Ising model as for the $\varphi^4$ model.  However, the functions ${\cal F}_c$ and ${\cal F}^{\rm{Is}}_c$ are expected to differ if they are plotted as functions of the coupling ratios $J_d/J$ and $E_3/E$, respectively. The physical origin of this nonuniversal effect is a bulk property arising from the different dependence of the  ratios of the principal  correlation lengths $q$ and $q^\text{Is}$ on the microscopic coupling ratios. More specifically,  Fig.~\ref{fig:q}  shows $q$ and $q^{\rm{Is}}$ as functions of $J_d/J$ and $E_3/E$, respectively. They differ for all values of the coupling ratios, except for two special cases, namely for $J_d=E_3 =0$, where $q=q^{\rm{Is}}=1$, and for $J_d/J=E_3/E=1$, where $q=q^{\rm{Is}}=\sqrt{3}$. While the former case corresponds to asymptotically isotropic $\varphi^4$  and Ising models with isotropic NN couplings on the same square lattice, the latter case corresponds to anisotropic systems on a square lattice with equal NN and NNN couplings in which case both models can be transformed to asymptotically isotropic systems with isotropic NN couplings on the same triangular lattice (for the $\varphi^4$ model see Fig. 2 (c) of \cite{dohm2006}, for the Ising model see \cite{salas}).
%
\begin{figure}[t!]
\begin{center}
\includegraphics[width=\columnwidth]{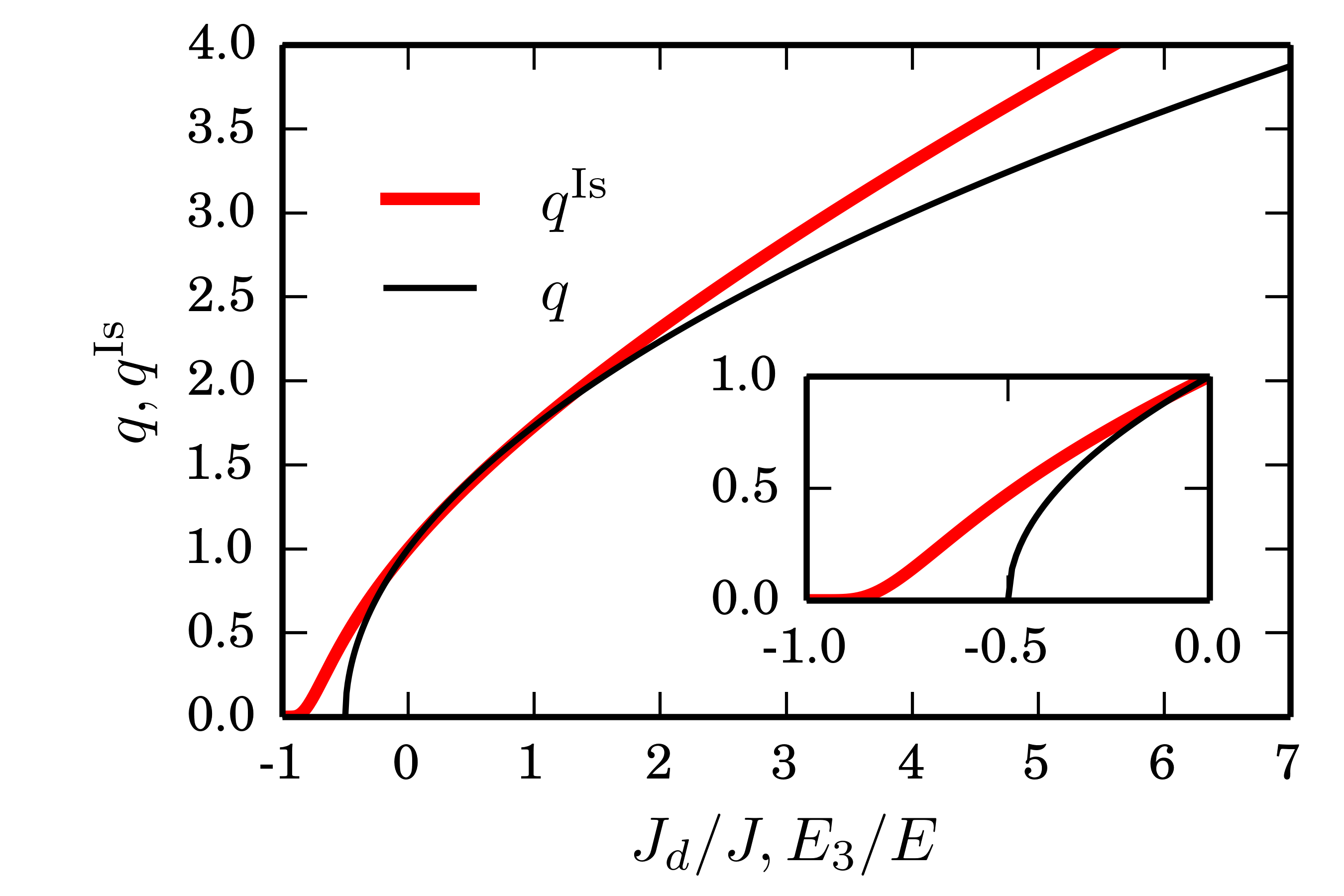}
\end{center}
\caption{
Nonuniversal dependence of the ratios $q$, (\ref{qratio}),  and $q^\text{Is}$, (\ref{ratioqising}),  of the asymptotic principal correlation lengths as functions of the coupling ratios $J_d/J$ and $E_3/E$ for the anisotropic $\varphi^4$ theory  and Ising model, respectively.
The inset focuses on the lower limit of the regime of weak anisotropy .
}
\label{fig:q}
\end{figure}

The above expectation is  confirmed by the nonuniversal plot of  ${\cal F}_c$ and ${\cal F}^{\rm{Is}}_c$ as functions of the ratios $J_d/J$ and $E_3/E$, shown in Fig.~\ref{fig:Fcomp}. Thus ${\cal F}_c$ and ${\cal F}^{\rm{Is}}_c$ depend, contrary to the isotropic case, on microscopic details.  This  illustrates the severe breakdown of two-scale-factor universality for anisotropic systems, and the fact that the knowledge of ${\cal F}_c$ or ${\cal F}^{\rm{Is}}_c$ for only the isotropic case has no predictive power for providing these quantities at any generic value of the microscopic couplings.  For this purpose, further (nonuniversal) knowledge of the  corresponding asymptotic principal correlation lengths ratio is mandatory (as well as of the orientation angle $\Omega$ of the principal axes in the general case~\cite{dohm-wessel-2021}).

\begin{figure}[t!]
\begin{center}
\includegraphics[width=\columnwidth]{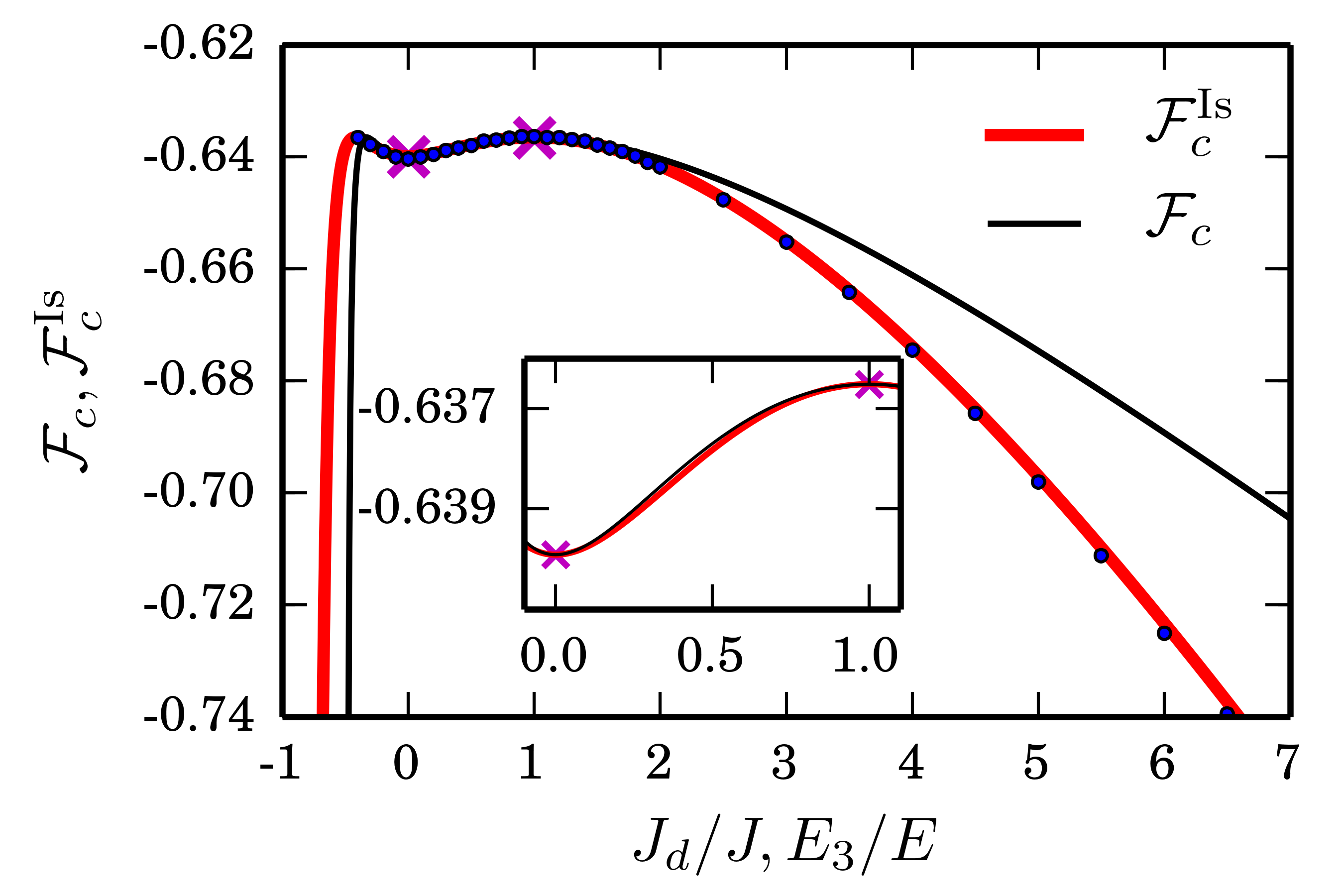}
\end{center}
\caption{
Comparison of the
exact analytic result  ${\cal F}_c$, (\ref{calFCFT}) for the anisotropic $\varphi^4$ model, (\ref{2a}),(\ref{matrixA}),
 and the prediction ${\cal F}_c^\text{Is}$, (\ref{calFCFTIs}) for the anisotropic Ising model, (\ref{IsingH}),
as functions of the coupling ratios $J_d/J$ and $E_3/E$, respectively. Also included
are the finite-size estimates ${\cal F}^\text{Is}_{c,L}$ from MC simulations for the anisotropic Ising model with $L=16$ (circles).
The statistical uncertainty of the MC data is smaller than the  symbol size.
The crosses at $J_d=E_3=0$ and $J_d/J=E_3/E=1$ are the analytical results from Refs.~\cite{FF} and  \cite{salas}, respectively.
The inset focuses on the crossing points of both curves for the special coupling ratios 0 and 1.
}
\label{fig:Fcomp}
\end{figure}
Nevertheless the feature of multiparameter universality ensures that the exact result shown in Fig.~\ref{fig:Fc}, based on CFT and $\varphi^4$ theory,
should be valid, as a function of the appropriate ratio $q$ of principal correlation lengths, for all weakly anisotropic systems of the $(d=2,n=1)$ universality class with the same principal axes in a square geometry with periodic BC. 
\renewcommand{\thesection}{\Roman{section}}
\renewcommand{\theequation}{6.\arabic{equation}}
\setcounter{equation}{0}
\section{ Critical Casimir amplitude}
We turn to the question as to the consequences of these results for the critical Casimir amplitude of the models analyzed above. For this purpose we need to extend the square geometry to a rectangular $L_\parallel \times L$  geometry with PBC where  the singular part ${\cal F}_c(\rho,q,\Omega)$ of the total free energy at $T_c$ now becomes a function of the  aspect ratio $\rho=L/L_\parallel$ \cite{dohm-wessel-2021}.
The explicit expression for ${\cal F}_c$ at $\Omega=\pi/4$ within the $\varphi^4$ model in terms of the modular parameter (\ref{tauq}) is \cite{dohm-wessel-2021}
\begin{eqnarray}
\label{}
{\cal F}_c(\rho,q,\pi/4)= {\cal F}^\mathrm{CFT}_c(\rho\: \tau[q])=-\ln{Z}^\mathrm{CFT}_c(\rho\: \tau[q]).
\end{eqnarray}
The amplitude $X_c(\rho,q,\pi/4)$ of the critical Casimir force in the vertical direction is obtained as \cite{dohm-wessel-2021}
\begin{eqnarray}
\label{Xcas}
X_c(\rho,q,\pi/4)= -\rho^2 \partial {\cal F}_c(\rho,q,\pi/4)/\partial \rho.
\end{eqnarray}
This yields
\begin{eqnarray}
{X}_c(\rho,q,\pi/4) = \rho^2 \frac{\partial {Z}^\mathrm{CFT}_c(\rho\: \tau[q])  /\partial \rho } {{Z}^\mathrm{CFT}_c(\rho\: \tau[q])} .
\end{eqnarray}
While the above results lead to a finite critical Casimir amplitude for any values of $\rho$ ~\cite{dohm-wessel-2021} in general,
we can prove that $X_c(\rho,q,\pi/4)$ vanishes for the special case of a square geometry, $\rho=1$, irrespectively of the value of $q$. Namely, as  shown in App.~\ref{appb}, the derivate $\partial {Z}^\mathrm{CFT}_c(\rho\: \tau[q])  /\partial \rho$
vanishes exactly at $\rho=1$, due to the properties of the Jacobi theta functions that enter ${Z}^\mathrm{CFT}$, such that
\begin{eqnarray}
X_c(1,q,\pi/4)= 0,
\end{eqnarray}
for all values of $q$ (compare Fig. 3 (a) of \cite{dohm-wessel-2021} for $\Omega=\pi/4$, see also \cite{hucht2011} for $q=1$).  Thus, for the $\varphi^4$ theory on a square geometry, the critical Casimir  amplitude vanishes for  all values of $J_d/J$ within the regime of weak anisotropy. The same result applies to the  critical Casimir amplitude of the anisotropic Ising model on the square geometry for all values of $E_3/E$ within the regime of weak anisotropy,  provided that multiparameter universality is valid also for $\rho\neq 1, \Omega=\pi/4$. The latter assumption is needed since the definition of ${X}_c$ contains the derivative with respect to $\rho$. More generally, multiparameter universality predicts the critical Casimir amplitude to vanish for all weakly anisotropic systems on a square geometry in the ($d=2$, $n=1$)  universality class if the orientation angle $\Omega$ of the principal axes equals $\pi/4$. For the analysis of $X_c$ for general  $q$, $\Omega$ and $\rho$ we refer to \cite{dohm-wessel-2021}.
\renewcommand{\thesection}{\Roman{section}}
\renewcommand{\theequation}{7.\arabic{equation}}
\setcounter{equation}{0}
\section{ Planar anisotropy in three dimensions}
It has been pointed out \cite{dohm-wessel-2021} that, in the presence of PBC, modular invariance plays an important role not only in the finite-size effects of weakly anisotropic Ising-like systems in two dimensions but more generally in the three-dimensional O$(n)$-symmetric $\varphi^4$ theory with planar anisotropies. Here we incorporate the two-dimensional anisotropy discussed above in the three-dimensional $\varphi^4$ model with a $L\times L\times L$ cubic geometry (see  the inset in Fig.~\ref{fig:3D}, see also Fig. 11 of \cite{dohm2008}) in the large-$n$ limit and compare the ensuing exact result for ${\cal F}_c$ with that presented above in two dimensions.

\begin{figure}[t!]
\begin{center}
\includegraphics[width=\columnwidth]{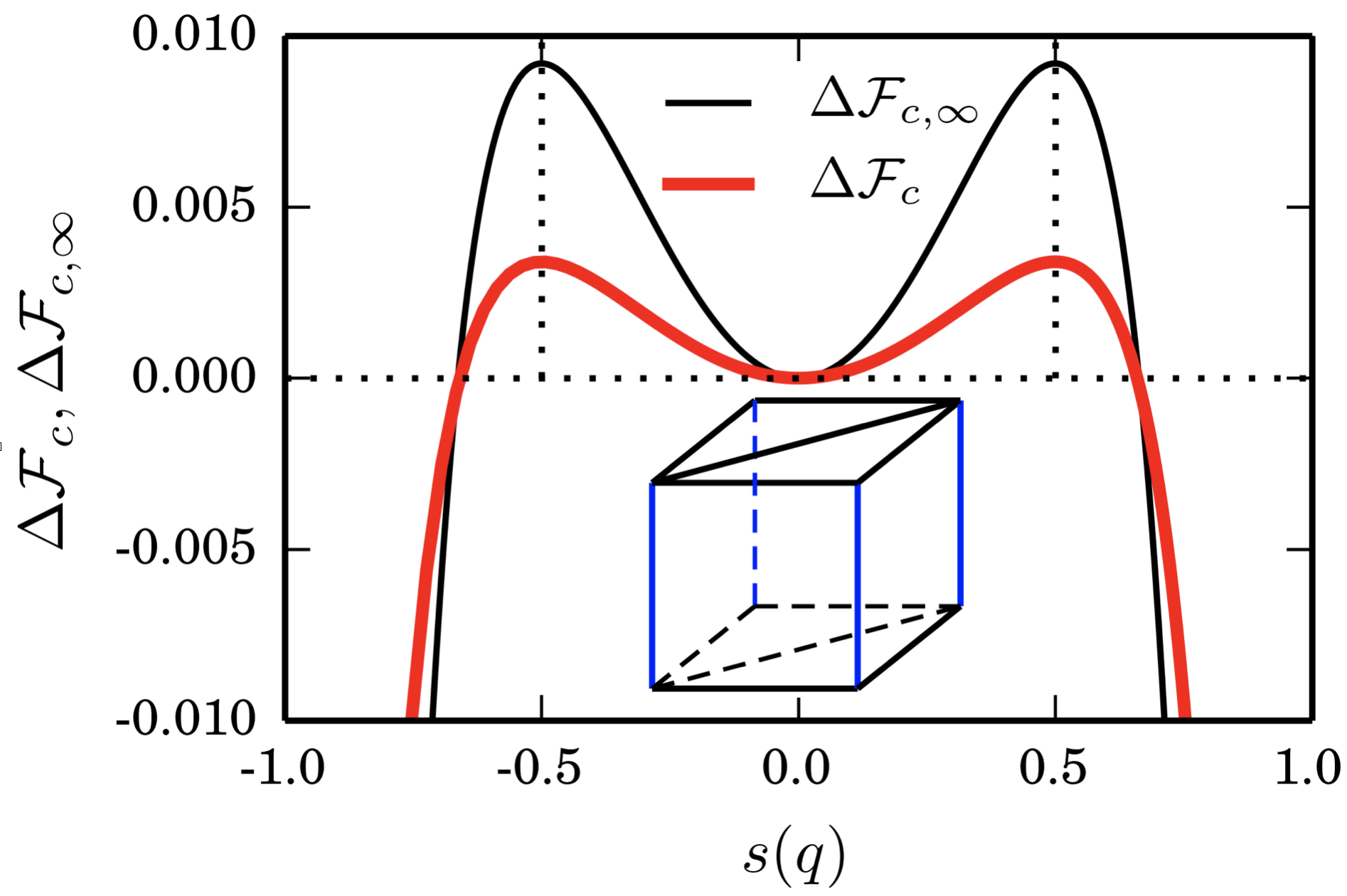}
\end{center}
\caption{
Exact analytic  result $ \Delta{\cal F} _{c,\infty}$, (\ref{DeltaFinfinity}), for the $d=3$ $\varphi^4$ theory at $T=T_c$ for $n=\infty$ shown as symmetric function of  $s(q)$, (\ref{sq1}), with maxima at $s(q)=\pm1/2$. For comparison, our exact CFT-based result $ \Delta{\cal F} _c,$ for the $d=2$, $n=1$ $\varphi^4$ theory in Fig.~\ref{fig:Fc} is also shown. The inset illustrates the unit cell of a simple-cubic lattice with a planar anisotropy as considered here.}
\label{fig:3D}
\end{figure}

Three-dimensional anisotropy and rescaling matrices of this $\varphi^4$ model can be chosen as (compare Eqs. (6.19) and (8.4) of \cite{dohm2018})
\begin{align}
\label{extmatrix3}
{\bf A}^{(x,y)}_3
=
\left(\begin{array}{ccc}
 {\bf A} & \hspace{0.15cm}{\bf 0} \vspace{0.15cm} \\
 {\bf  0^T} & \hspace{0.15cm}(\det{\bf A})^{1/2} \\
\end{array}\right),
{{\mbox {\boldmath$ \lambda$}}} = \left(\begin{array}{ccc}
   \lambda_1 & 0 & 0 \\
  0 &  \lambda_2 & 0 \\
  0 & 0 &\hspace{0.15cm}(\lambda_1\lambda_2)^{1/2}\\
\end{array}\right) \;
\end{align}
where ${\bm 0}= \left(\begin{array}{c}
  0 \\
  0  \\
\end{array}\right)$ and ${\bf 0^T}=(0\;\;0)$ and where ${\bf A}$ is the $2\times 2$ matrix (\ref{matrixA}) describing the anisotropy in the horizontal $(x,y)$ plane. The three eigenvalues are
$\lambda_1=2(J+2J_d),\lambda_2=2J, \lambda_3= (\lambda_1\lambda_2)^{1/2}=(\det{\bf A})^{1/2}$.
%
%

%
%

%
%
%
The reduced matrices  $\bar {\bf A}^{(x,y)}_3= {\bf A}^{(x,y)}_3/\big(\det{\bf A}^{(x,y)}_3\big)^{1/3}$ and ${{\mbox {\boldmath$\bar \lambda$}}}={{\mbox {\boldmath$ \lambda$}}}/(\det {\mbox {\boldmath$ \lambda$}})^{1/3}$  are
\begin{align}
\label{barextmatrix3}
{\bf \bar A}^{(x,y)}_3(q)
=
\left(\begin{array}{ccc}
 {\bf \bar A}(q) & \hspace{0.15cm}{\bf 0} \vspace{0.15cm} \\
 {\bf  0^T} & \hspace{0.15cm}1 \\
\end{array}\right), \;\;\;\;\;
{{\mbox {\boldmath$\bar \lambda$}}}(q) = \left(\begin{array}{ccc}
  \bar \lambda_1 & 0 & 0 \\
  0 & \bar \lambda_2 & 0 \\
  0 & 0 & 1\\
\end{array}\right)
\end{align}
with  ${\bf \bar A}(q)$ given by (\ref{Abarx}) and with $\bar \lambda_1=(\lambda_1/\lambda_2)^{1/2}=q=\xi_{0 +}^{(1)}/\xi_{0 +}^{(2)}$ and  $\bar \lambda_2=q^{-1}$ where $\xi_{0 +}^{(\alpha)}$ are the planar principal correlation-length amplitudes above $T_c$. This represents a three-dimensional $\varphi^4$ model with the same planar anisotropy as for the $d=2$ models discussed above, but with no anisotropy
of the bulk correlation function at $T_c$
$G_{b,c}({\bf x})\propto (... +x_3^2 )^{-(1+\eta)/2}$ \cite{dohm2018}
in the $z$-direction, owing to the particular choice of the principal correlation length
%
\be
\label{corr3}
\xi_{0 +}^{(3)}=\big[\xi_{0 +}^{(1)}\xi_{0 +}^{(2)}\big]^{1/2}
\ee
determined by the mean correlation length in the plane.  Substituting our matrix ${\bf \bar A}^{(x,y)}_3(q)$, (\ref{barextmatrix3}), into Eq. (6.39) of \cite{dohm2008} (see also the expressions in \cite{SM}, Sec. IV for
$\rho=1$) yields the exact critical amplitude ${\cal F}_{c,\infty}(q)=\lim_{n\to \infty}{\cal F}_c(q)/n$ in the large-$n$ limit for PBC which is shown in Fig.~\ref{fig:3D} as
\be
\label{DeltaFinfinity}
\Delta {\cal F}_{c,\infty}(q)= {\cal F}_{c,\infty}(q)-{\cal F}_{c,\infty}(1),
\ee
i.e.,  relative to the isotropic case $q=1$, with ${\cal F}_{c,\infty}(1)= -0.525524$, as a function of $s(q)$, (\ref{sq1}). The corresponding quantity $\Delta {\cal F}_{c}(q) = {\cal F}_{c}(q)-{\cal F}_{c}(1)$ is also shown for our exact CFT-based result for the $d=2$, $n=1$ case of Fig.~\ref{fig:Fc}. In the large-$n$ limit the anisotropy enters  ${\cal F}_{c,\infty}(q)$ through the function $K_3$ defined in Eq. (17) of \cite{dohm-wessel-2021} which displays the property of modular invariance parallel to that of $Z^{\text{CFT}}$. This explains the symmetry of the $n=\infty$ curve in Fig.~\ref{fig:3D}. In particular we again find a two-peak structure with equal-height maxima, with the same positions $ q_{\text{max}}= 3^{\pm 1/2}$ of the maxima  as for the anisotropic $d=2,n=1$ model discussed above. The symmetry with respect to $s$ was also found in \cite{dohm2008} where, however, the origin from modular invariance was not yet recognized.

The same symmetry persists at finite $n$, as we have verified on the basis of the approximate results of \cite{dohm2018} for the $d=3$ O$(n)$-symmetric $\varphi^4$ theory with the planar anisotropy defined above. Invoking multiparameter universality for general $1\leq n\leq \infty$ we predict the same results for three-dimensional $O(n)$-symmetric fixed-length spin models, after substituting $q \to q^\mathrm{spin}$, e. g., for $XY$ models $(n=2)$ or Heisenberg models $(n=3)$.

The symmetry with respect to $s(q)$, or, equivalently, with respect to $\ln q$, was misinterpreted as a signature of universality (rather than as a consequence of modular invariance) in the analysis \cite{kastening2013} of Monte Carlo data \cite{selke2005} for the critical Binder cumulant of the anisotropic $d=2$ Ising model. 
The analytic result of the anisotropic $d=3$ $\varphi^4$ theory for the Binder cumulant \cite{dohm2008} was not analyzed in terms of the correlation length ratio $q = (1+2J_d/J)^{1/2}$ and the violation of two-scale-factor universality arising from the nonuniversal coupling dependence of $q$ and $q^{\rm Is}$ shown in our Fig. 6 was not recognized in \cite{kastening2013}. See also the comments in \cite{dohm2018}.

\appendix

\section{Jacobi theta functions from \cite{franc1997}}
\label{appa}
\renewcommand{\theequation}{A.\arabic{equation}}\setcounter{equation}{0}
We follow \cite{franc1997} for the notation of   the  Jacobi theta functions $\theta_i(\tau)$ used in (\ref{ZIsing}). For  a complex number  $\tau$ with
$\mathrm{Im}({\tau})>0$ they are denoted by
\begin{eqnarray}
\theta_2(\tau)&=&\sum_{n=-\infty}^{\infty} q^{(n+1/2)^2/2},\label{SMtf2}\\
\theta_3(\tau)&=&\sum_{n=-\infty}^{\infty} q^{n^2/2},\label{SMtf3}\\
\theta_4(\tau)&=&\sum_{n=-\infty}^{\infty} (-1)^nq^{n^2/2},\label{SMtf4}
\end{eqnarray}
with $q=\exp(2\pi i \tau)$.
\section{Proof of $\partial {Z}^\mathrm{CFT}_c(\rho\: \tau[q])  /\partial \rho|_{\rho=1}=0$ }
\label{appb}
\renewcommand{\theequation}{B.\arabic{equation}}\setcounter{equation}{0}
Besides the modular invariance of ${Z}^\mathrm{CFT}_c(\tau)$, which gives
\begin{equation}
{Z}^\mathrm{CFT}_c(\tau)={Z}^\mathrm{CFT}_c(-1/\tau)
\end{equation}
we obtain the following property of ${Z}^\mathrm{CFT}_c(\tau)$ with respect
to complex conjugation of $\tau$,
\begin{equation}\label{appb2id1}
{Z}^\mathrm{CFT}_c(\tau)={Z}^\mathrm{CFT}_c(-\tau^*)
\end{equation}
from the fact that $|\theta_i(\tau)|=|\theta_i(-\tau^*)|$, for $i=2,3,4$, which follows
directly from the series representations given in App.~\ref{appa}.
Furthermore, since for the considered $\varphi^4$ theory and Ising model
the modular parameter has unit length, $|\tau[q]|=1$, we get
\begin{eqnarray}
\label{appb2id2}
-1/\tau[q]=-\tau[q]^*.
\end{eqnarray}
We next calculate
$\partial {Z}^\mathrm{CFT}_c(\rho\: \tau[q])  /\partial \rho|_{\rho=1}$
as
\begin{eqnarray}
\label{appblim}
\lim_{\epsilon\rightarrow 0}
\frac{  {Z}^\mathrm{CFT}_c((1+\epsilon)  \tau[q]) -  {Z}^\mathrm{CFT}_c((1-\epsilon)  \tau[q])}{2 \epsilon}.
\end{eqnarray}
Due to modular invariance,
\begin{eqnarray}
{Z}^\mathrm{CFT}_c((1+\epsilon)  \tau[q])& =& {Z}^\mathrm{CFT}_c(-1/((1+\epsilon)  \tau[q]))\nonumber \\
&=&{Z}^\mathrm{CFT}_c((1-\epsilon)\: (-1/ \tau[q]))+O(\epsilon^2).\nonumber
\end{eqnarray}
Now using the identities (\ref{appb2id1}) and (\ref{appb2id2}), we obtain:
\begin{eqnarray}
{Z}^\mathrm{CFT}_c((1-\epsilon)(-1/ \tau[q]))
&=&{Z}^\mathrm{CFT}_c((1-\epsilon)(- \tau[q]^*))\nonumber\\
&=&{Z}^\mathrm{CFT}_c((1-\epsilon) \tau[q])),\nonumber
\end{eqnarray}
such that we find:
\begin{eqnarray}
{Z}^\mathrm{CFT}_c((1+\epsilon) \tau[q]))={Z}^\mathrm{CFT}_c((1-\epsilon) \tau[q]))+O(\epsilon^2).\nonumber
\end{eqnarray}
Using this in (\ref{appblim}), we thus find that
\begin{eqnarray}
\partial {Z}^\mathrm{CFT}_c(\rho\: \tau[q])  /\partial \rho|_{\rho=1}=0.
\end{eqnarray}

\end{document}